# Disability, Job Satisfaction, and Workplace Accommodations: Evidence from the Healthcare Industry

*Yana van der Meulen Rodgers, Rutgers University, New Brunswick, NJ, USA*

*Lisa Schur, Rutgers University, New Brunswick, NJ, USA*

*Flora M. Hammond, Indiana University School of Medicine, Indianapolis, IN, USA*

*Renee Edwards, Rutgers University, New Brunswick, NJ, USA*

*Jennifer Cohen, Rutgers University, New Brunswick, NJ, USA*

*Douglas Kruse, Rutgers University, New Brunswick, NJ, USA*

**Abstract:**
**Purpose.** This paper examines the extent to which job satisfaction, requests for accommodations, and the likelihood of a request being granted vary by disability status. We further analyze whether being granted workplace accommodations moderates the relationship between work satisfaction and disability.
**Methods.** We use a novel survey of healthcare workers centered on disability status, perceptions of work experiences, and the provision of accommodations. The data are used in a descriptive analysis and multiple regressions to examine the moderating effect of accommodations on the relationship between disability and indicators related to job satisfaction.
**Results.** Results show that people with disabilities have more negative perceptions of their work experiences than people without disabilities. Although people with disabilities are more likely to request accommodations than people without disabilities, they are equally likely to have their requests wholly or partly granted. Regression results indicate that the negative relationships between disability status and most measures of work experience are largely eliminated when accounting for the disposition of accommodation requests. The main exception is turnover intentions, in which the adverse relationship with having a disability does not change even when an accommodation is granted. Partly granting accommodations is helpful only for some metrics of job experience.
**Conclusion.** Our paper shows that fully granting accommodations can go a long way to closing the disability gap in job satisfaction between people with and without disabilities.

**Corresponding Author:** Yana van der Meulen Rodgers, School of Management and Labor Relations, 94 Rockafeller Road, Rutgers University, New Brunswick, NJ 08544. Email yana.rodgers@rutgers.edu.

**Acknowledgments:** The authors acknowledge So Ri Park for her valuable research assistance. We also thank Mason Ameri, Lata Gangadharan, Josefina Tranfa-Abboud, Michael O'Hara, Stephanie Rizzardi, Art Eubank, David Macpherson, the West Virginia University Center for Excellence in Disabilities Seminar, the National Association for Forensic Economics session at the 2025 ASSA conference, the Doctorado de Ciencias Sociales at Universidad de Buenos Aires, and two anonymous reviewers for their helpful suggestions. This research received funding support from the National Institute on Disability, Independent Living, and Rehabilitation Research (NIDILRR) for the Rehabilitation Research & Training Center (RRTC) on Employment Policy: Center for Disability-Inclusive Employment Policy Research Grant [grant number #90RTEM0006-01–00] and the NIDILRR RRTC on Employer Practices Leading to Successful Employment Outcomes Among People with Disabilities Research Grant [grant number #90RTEM0008-01-00].

**Statements and Declarations:**

**Competing Interests.** The authors declare no potential conflicts of interest with respect to the research, authorship, and/or publication of this article.

# 1. Introduction

The literature clearly shows disparities in job satisfaction between people with and without disabilities [1]. These disparities partly reflect the psychological toll of stigma and the threat that stigma poses to people's social identities. Employees with disabilities often face subtle forms of bias and exclusion that influence their experiences on the job [2, 3]. Stigma can trigger identity threats, leading individuals to engage in self-regulation strategies that may protect their status but also incur hidden psychological costs [4]. Reducing identity threat through supportive environments that include workplace accommodations may help to improve job satisfaction and perceptions of employees' work experiences that contribute to job satisfaction. Yet employees with disabilities often navigate complex decisions around requesting accommodations, balancing the potential benefits of disclosure against the risks of bias and marginalization [5]. Although accommodation requests are a common reason for disability disclosure, individuals may also disclose disabilities to foster openness, build trust with supervisors, or explain certain behaviors or needs [6].

Disability remains one of the most underrepresented dimensions in organizational diversity and inclusion efforts, despite its prevalence in the workforce. While many organizations have adopted diversity initiatives, disability is often excluded from these frameworks, resulting in persistent disparities in workplace experiences and labor market outcomes such as earnings and employment rates [7, 8]. Meaningful disability inclusion requires systemic changes in human resource management, leadership accountability, and organizational culture [6]. In this context, inclusion is not merely about physical access or compliance with legal standards but about fostering workplace environments where employees with disabilities can fully participate and thrive [9]. This perspective aligns with broader calls in organizational psychology to move beyond performative inclusion and toward structural equity that addresses attitudinal and institutional barriers [10].

In this context, our study uses a novel dataset from the healthcare industry to analyze how often and to what degree workers are granted accommodations, how work satisfaction measures vary between people with and without disabilities, and the extent to which workplace accommodations moderate the relationship between work satisfaction and disability. The potential for accommodations to improve job satisfaction for people with disabilities has been explored before [11], but not in such detail, compared with those without disabilities, or in the context of pandemic-related changes in thinking about remote work and scheduling flexibility. Our data are based on a survey centered on the perceptions of people's experiences at work and the provision of accommodations in the workplace. We compare the perspectives of workers with and without disabilities, paying particular attention to their likelihood of requesting and being granted accommodations to help them do their job better.

Our measures of perceptions of work experiences are based on a direct survey question about job satisfaction and ten indicators based on scales in the human resource management and organizational behavior literature, including social exchange theory. According to this theoretical construct, individuals evaluate their relationships with organizations based on perceived reciprocity and fairness [12, 13]. When organizations respond positively to accommodation requests or demonstrate inclusive practices, employees are more likely to feel valued and reciprocate with positive attitudes and behaviors. Conversely, when support is lacking or inconsistent, employees may perceive an imbalance in the exchange relationship, leading to disengagement or reduced organizational commitment. This framework highlights the importance of relational trust and perceived organizational support in shaping the work experiences of people with disabilities.

We examine the healthcare industry for several reasons. First, quality of care, and therefore the health of the population, is largely determined by the resources allocated to and the capacity of the healthcare system. Second, while the overall prevalence of disability among healthcare practitioners is close to the average for all U.S. workers, support and service roles within healthcare—which often include lower-wage, frontline positions—have substantially higher disability rates [14]. Third, studying the healthcare industry allows us to examine a diverse range of employees since underrepresented men and women of all races and ages disproportionately work in the sector, and the wide range of health occupations requires workers with all levels of education. Finally, employers invested (in theory) in generating wellbeing may be more willing to accommodate workers with disabilities, thus making the healthcare industry a particularly relevant sector to examine.

We hypothesize that (1) the proportion of people who request accommodation will be higher among people with a disability; (2) people with disabilities will have lower job satisfaction and more negative perceptions of work experiences compared to people without disabilities; and (3) being granted accommodations will have a positive moderating effect on the relationship between disability and measures of work experiences.

**2. Materials and Methods**

We partnered with a major statewide university-based health system to conduct a novel survey of healthcare workers with and without disabilities, assessing their experiences with employer policies related to accommodations and various measures of job satisfaction. Our research team worked with collaborators in this system to distribute to employees a link to our online survey in Qualtrics, along with a cover note about the study and information about informed consent. Employees were notified about the survey in the organization's online newsletter. More specifically, our survey description and link were communicated as a "story" on the employer's team portal, and the team portal story was included in the daily newsletter email distributed to all employees (approximately 36,000). Direct emails to all employees with our survey invitation and link were not permitted. A total of 1,405 employees in the organization took the survey between May 26 and July 31, 2023. We cannot calculate a response rate because we do not know how many employees read the newsletter. After dropping observations with missing values, we are left with a sample size of 993 respondents.

Our survey instrument included questions on accommodations requested, whether requests were granted/employer response to requests, and employees' awareness and perceptions of employer policies that address workers' physical and mental health needs. The design of the survey instrument was guided by survey questions highlighted in Schur et al. [11, 15] and includes scales commonly found in the organizational behavior literature. These scales are discussed in the notes section of the online Appendix.

We used the data to calculate simple summary statistics on disclosure of disability to the employer, prevalence of accommodation requests by disability status, employer responsiveness to such requests, perceptions of workplace inclusiveness, perceptions of the treatment of people with disabilities, and various measures of job satisfaction. Along with a descriptive analysis of sample means, the data are used in multiple regressions to examine whether being granted accommodations moderates the relationship between disability and indicators of work experiences, controlling for other characteristics. The complete model specification is:

$$Outcome_i = b_0 + b_1 Disab_i + b_2 Accomm_i + b_3 Disab_i * Accomm_i + b_4 X_i + e_i \ .$$

The notation *Outcome*$_i$ is job satisfaction and a vector of 10 indicators of perceptions of work experiences for person i, *Disab*$_i$ denotes a dummy variable for disability status, and *Accomm*$_i$ is an indicator for the disposition of accommodation requests: denied, partly granted, and fully granted, with "no accommodation request" serving as the base category. The notation $X_i$ is a set of demographic characteristics including age, gender, race/ethnicity, marital status, education, income above $75,000, number of children at home, occupation, full-time worker, and tenure at the employer. Job satisfaction is measured directly by a survey question and workplace experiences are measured by ten indices are constructed from survey questions about job autonomy [16-18], turnover intentions [19], organizational commitment [20], organizational citizenship behaviors [21], perceived organizational support [22, 23], employer openness to differences, the climate for inclusion [24], treatment of people with disabilities, relationship with one's manager (also known as leader-member exchange) [25], and relationships with one's coworkers (also known as coworker exchange) [26]. These outcomes are discussed further in the next section. In both the descriptive analysis (the 2-tail t tests for all disability gaps) and the regression analysis we report statistical significance as *** ($p<0.01$), ** ($p<0.05$), and *($p<0.10$), where the results with $p<0.10$ are considered to be marginally significant or approaching statistical significance.

We measure disability using the six Census questions, supplemented by two additional questions on difficulty with social interactions and long-term activity limitations. The eight questions are: (1) "Are you deaf or do you have serious difficulty hearing?"; (2) "Are you blind or do you have serious difficulty seeing even when wearing glasses?"; (3) "Do you have serious difficulty concentrating, remembering, or making decisions?"; (4) "Do you have serious difficulty walking or climbing stairs?"; (5) "Do you have difficulty dressing or bathing?"; (6)

"Do you have difficulty doing errands alone such as visiting a doctor's office or shopping?"; (7) "Do you have difficulty interacting and/or communicating with others?"; and (8) "Do you have a long-term health problem or impairment that limits the kind or amount of work, housework, school, parenting, recreation, or other activities you can do?" A person who answers yes to any of these questions is defined as having a disability.

Because disclosure gives employers an opportunity to accommodate employee needs, our analyses use two alternative constructions of the disability sample: (1) those who reported a disability in the survey (n=228), and (2) those who have disclosed a disability to their employers (n=114). The construction of these alternative disability samples and the relevant comparison groups is depicted in Figure 1. In both cases, the total sample is 993 individuals. To limit the number and size of our tables, results based on self-reported disability status are reported in the main tables, and results based on disclosed disability status are reported in Appendix Tables 2-4.

Figure 1. Construction of Disability Subsamples and Comparison Groups

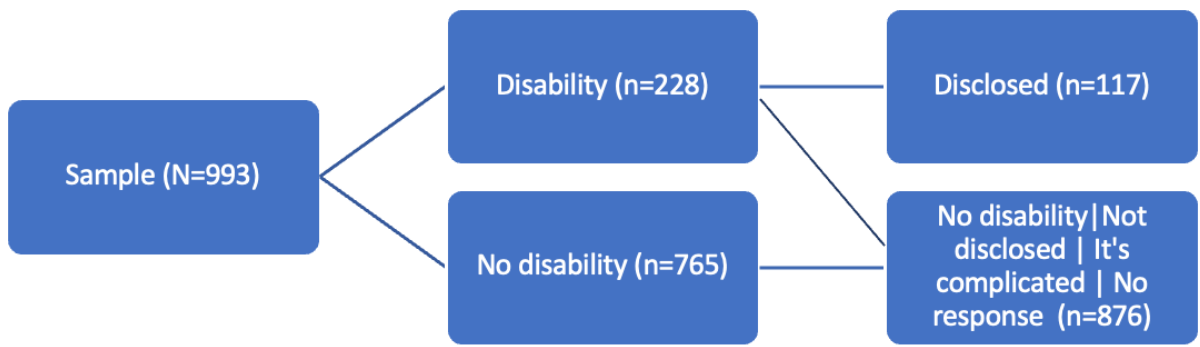

## 3. Sample Characteristics

### 3.1 Prevalence and types of disabilities

Table 1 shows that of the 993 respondents in our sample, 23.0% report functional limitations or challenges in social interactions. We refer to this response as having a disability or health impairment. This percentage drops to 18.4% if we restrict our measure to the six kinds of disability reported in Census data, almost double the 9.4% estimate based on data from the 2024 American Community Survey (ACS) for the percentage of healthcare workers with a disability in the organization's state. One explanation is that we may have over-sampled people with disabilities, and another possibility is that the employer where we conducted our survey has a relatively higher level of employees with disabilities compared to the population of healthcare workers in the state at large. Nonetheless, we have sufficient sample sizes of people living with at least one disability and people without a disability to conduct comparative analyses. Table 1 further shows that among people with a disability, close to half (46.5%) reported a long-term impairment. The most common type of impairment was difficulty concentrating and making decisions, followed by difficulty walking and climbing stairs.

Table 1. Sample Statistics on Disability Status

|  | *Number of respondents* | *Percent of respondents* |
|---|---|---|
| *Total Sample* | 993 | 100.0 |
| Person without disability | 765 | 77.0 |
| Person with disability | 228 | 23.0 |
|  |  |  |
| *Types of Disabilities (Not Mutually Exclusive)* | 228 | 100.0 |
| Deaf/difficulty hearing | 33 | 14.5 |
| Blind/difficulty seeing | 10 | 4.4 |
| Difficulty concentrating/making decisions | 113 | 49.6 |
| Difficulty walking/climbing stairs | 57 | 25.0 |
| Difficulty dressing/bathing | 7 | 3.1 |
| Difficulty doing errands alone | 39 | 17.2 |
| Difficulty interacting with others | 47 | 20.6 |
| Long-term health impairment | 106 | 46.5 |
| *Difficulties Among People with Disabilities* | 228 | 100.0 |
| Health condition has affected my ability to complete work duties with moderate or severe difficulty | 32 | 14.0 |
| Have you disclosed your health condition, impairment, or disability to your employer? | | |
|   Yes | 117 | 51.8 |
|   No | 57 | 25.2 |
|   It's complicated | 52 | 23.0 |
|   Did not respond | 2 | 0.9 |
| I have sometimes been unfairly treated because of my health condition, impairment, or disability. | 34 | 14.9 |
| At work I feel socially isolated because of my health condition, impairment, or disability | 35 | 15.4 |
| I have not disclosed my health condition, impairment, or disability at work because I am afraid of being stigmatized | 38 | 16.7 |

Source: Authors' computations based on original survey.

Of the group with a disability, about one-half (51.8%) disclosed their disability to their employer, another quarter (25.2%) did not disclose, and the remainder said it's complicated (23.0%) or did not respond (0.9%). Applied to the entire employee population, this indicates that about 12% (0.518*0.23) of all surveyed employees said they disclosed a disability to their employer, which is greater than the national average of 4.2% [27]. The higher disclosure rate observed in our sample of healthcare workers could be attributed to several factors, such as the demanding nature of their roles, which may necessitate accommodations; their increased familiarity with disability-related legislation; the inclusive workplace policies implemented by this specific employer; and the prevalence of disabilities related to Long Covid given the recent pandemic, a mass disabling event.

The sample provides us with a valuable opportunity to examine differences between people who have and have not disclosed at the same employer. As shown in Appendix Table 1, the likelihood of disclosure varies somewhat by type of disability. These conditional probabilities indicate that people with difficulty walking and/or climbing stairs are most likely to disclose their disability to the employer (68.4%), followed by people with long-term health impairments (61.3%). Recall that these disability categories are not mutually exclusive. People who have disabilities considered to be invisible (difficulty concentrating/making decisions, difficulty interacting with others) have lower rates of disclosure at 40.7% and 40.4%, respectively. These two groups are most likely to say they did not disclose to the employer (as opposed to responding that "it's complicated"). To the extent that the pandemic and Long Covid may have reduced stigma toward disability and increased the likelihood of disclosure, we do not appear to be capturing this effect among people with difficulty concentrating/making decisions. Still, we may be capturing this effect among people with long-term health impairments.

Of the employees with a disability, 14.0% indicated experiencing moderate or severe challenges in performing their work duties, 46.9% reported only minor difficulties, and 39.0% stated they had no difficulties. Moreover, 14.9% and 15.4% of respondents with disabilities reported some form of mistreatment or isolation at work, respectively. A closer look indicates that about 60% of these respondents have disclosed their disability to the employer, suggesting that disclosure appears to be associated with experiencing mistreatment or isolation. The final row of Table 1 shows that 16.7% of respondents with disabilities have not disclosed their health conditions due to fear of stigma at work, which seems low if disclosure is associated with mistreatment or isolation.

**3.2 Demographic characteristics**

Sample means for people with and without disabilities show meaningful differences in the number of children, education, marital status, and income (Table 2). Individuals with disabilities are less likely to have a college degree and be married compared to people without disabilities. People with disabilities also have fewer children and lower incomes on average compared to people without disabilities. These findings on key demographic indicators among people with and without disabilities are consistent with previous studies [28]. Sample means further show that workers with disabilities were less likely to have worked at the organization before the pandemic and were more likely to have relatively short tenures (five years or less). However, there are no statistically significant differences between people with and without disabilities in the likelihood of working full-time or having a management role. Among the eight occupational categories in the survey, the most common occupation among respondents with and without disabilities is "other healthcare provider" (which includes jobs such as phlebotomist and pharmacist), followed closely by administrative support staff. The largest disability gap among the occupational categories is for nurses: people with disabilities are considerably less likely to be nurses (18.5%) compared to people without disabilities (26.7%). Overall, the results for people with disclosed disabilities are comparable (Appendix Table 2), except that the differences by disability status in being multiracial/other race and having a professional degree or PhD are no longer statistically significant.

Table 2. Sample Means for Demographic Characteristics by Disability Status (Measured as Proportions Unless Indicated Otherwise)

| Demographic Characteristic | *Disability Mean* | *No Disability Mean* | *Difference* |
|---|---|---|---|
| Age | 43.638 | 44.935 | -1.297 |
| # Children at home | 1.772 | 2.050 | -0.278*** |
| Gender | | | |
|   Man | 0.085 | 0.108 | -0.023 |
|   Woman | 0.848 | 0.868 | -0.020 |
|   Nonbinary | 0.067 | 0.024 | 0.043*** |
| Race/Ethnicity | | | |
|   Black | 0.124 | 0.118 | 0.007 |
|   White | 0.836 | 0.837 | -0.002 |
|   Hispanic | 0.040 | 0.026 | 0.014 |
| American Indian/Alaska Native | 0.058 | 0.034 | 0.023 |
| Asian/Pacific Islander | 0.013 | 0.017 | -0.004 |
| Multiracial/other | 0.089 | 0.052 | 0.037** |
| Married | 0.482 | 0.666 | -0.184*** |
| Education |  |  |  |
| <9th grade | 0.004 | 0.000 | 0.004* |
| High school graduate | 0.098 | 0.069 | 0.029 |
| Some college, no degree | 0.209 | 0.147 | 0.062** |
| Associate degree | 0.204 | 0.167 | 0.038 |
| Bachelor's degree | 0.311 | 0.371 | -0.060 |
| Master's degree | 0.151 | 0.181 | -0.030 |
| Professional degree/PhD | 0.022 | 0.065 | -0.043** |
| Income >=$75,000 | 0.223 | 0.364 | -0.141*** |
| Works full-time | 0.886 | 0.882 | 0.004 |
| Worked at employer before pandemic | 0.741 | 0.800 | -0.059* |
| Worked at employer <= 5 years | 0.513 | 0.400 | 0.113*** |
| Occupation |  |  |  |
| Administrative support staff | 0.194 | 0.176 | 0.017 |
| Professional/technical staff | 0.167 | 0.122 | 0.046* |
| Nurse | 0.185 | 0.267 | -0.082** |
| Physician | 0.000 | 0.008 | -0.008 |
| Healthcare aide | 0.079 | 0.041 | 0.039** |
| Other healthcare provider | 0.225 | 0.209 | 0.016 |
| Manager | 0.088 | 0.125 | -0.037 |
| Other service provider | 0.062 | 0.052 | 0.009 |

Note: Sample size 993. *** statistically significant at 1%, ** at 5%, and * at 10% in 2-tail t tests. Results denote the proportion of respondents who gave the indicated response in the survey, ranging from 0 to 1. See Appendix Table 2 for sample means using disclosed disability status.

## 4. Accommodation Requests Among Employees With and Without Disabilities

Table 3 shows that people with and without disabilities have requested accommodations to help them do their jobs better: 70.2% of people with disabilities have requested accommodations, compared to 56.0% of people without disabilities. Respondents are particularly likely to request changes in work schedules, indicating that flexibility is a major issue for all workers. Still, the likelihood of requesting a schedule change is even larger for people with disabilities (65.8% and 55.4%, p<0.05). Some of these accommodation requests may be fairly minor (e.g., starting work at 9:00 am instead of 8:30 am to permit time for getting children ready for school), thus explaining the high percentage of people with and without disabilities requesting accommodations. Another common request is a change in communications and information sharing, and people with disabilities are considerably more likely to make this request (55.7% and 43.7%, p<0.01).

Table 3. Sample Means for Accommodation Requests by Disability Status (Measured as Proportions)

| Accommodations | Disability Mean | No Disability Mean | Difference |
|---|---|---|---|
| Have you ever requested accommodations? | 0.702 | 0.560 | 0.142*** |
| Type requested: equipment | 0.443 | 0.484 | -0.041 |
| Type requested: physical change to workplace | 0.297 | 0.256 | 0.042 |
| Type requested: work from home | 0.297 | 0.275 | 0.023 |
| Type requested: change to work schedule | 0.658 | 0.554 | 0.104** |
| Type requested: restructure job | 0.241 | 0.195 | 0.046 |
| Type requested: move to another job or location | 0.203 | 0.143 | 0.059* |
| Type requested: change communications/info sharing | 0.557 | 0.437 | 0.120*** |
| Type requested: other | 0.241 | 0.181 | 0.060 |
| Most recent request: equipment | 0.161 | 0.218 | -0.056 |
| Most recent request: physical change to workplace | 0.050 | 0.061 | -0.011 |
| Most recent request: work from home | 0.087 | 0.059 | 0.028 |
| Most recent request: change to work schedule | 0.273 | 0.288 | -0.015 |
| Most recent request: restructure job | 0.099 | 0.075 | 0.024 |
| Most recent request: move to another job or location | 0.075 | 0.056 | 0.018 |
| Most recent request: change communications/info sharing | 0.168 | 0.176 | -0.008 |
| Most recent request: other | 0.087 | 0.068 | 0.019 |
| Most recent accommodation was requested within the past 12 months | 0.791 | 0.831 | -0.040 |
| Did you request this change in order to accommodate any health condition, impairment, or disability that you may have? | 0.331 | 0.077 | 0.254*** |
| Was the requested change or accommodation made? | | | |
|   Yes | 0.500 | 0.475 | 0.025 |
|   No | 0.259 | 0.304 | -0.044 |
|   Partially | 0.241 | 0.221 | 0.019 |

Note: Sample size 993 for first question. Responses for remaining questions are conditional on having ever requested accommodations. Results denote the proportion of respondents who gave the indicated response in the survey, ranging from 0 to 1. See Appendix Table 3 for sample means using disclosed disability status.
*** statistically significant at 1%, ** at 5%, and * at 10% in 2-tail t tests.

Note that the types of accommodations listed in Table 3 are consistent with the Americans with Disabilities Act (ADA), which defines reasonable accommodations as "modifications or adjustments to a job, the work environment, or the way things are usually done during the hiring process that enable a qualified individual with a disability to enjoy an equal employment opportunity" [29]. According to the ADA, reasonable accommodations can include part-time or modified work schedules, job restructuring, and acquisition or modification of equipment or devices. Table 3 further shows that although the majority of people with and without disabilities have requested accommodations, only 7.7% of those without disabilities requested the change in order to accommodate a health condition, impairment, or disability (as opposed to some other reason, like caring responsibilities) compared to 33.1% of those with disabilities. Despite these gaps in health needs, there is no statistically significant difference between workers with and without a disability in the likelihood of having an accommodation request being granted. Only half of all workers have their accommodation requests fully granted, and another 22% to 24% have their requests partially granted.

**5. Job satisfaction and perceptions of work among people with and without disabilities**

Survey results provide compelling evidence of negative workplace experiences for people with disabilities. In particular, Table 4 reports sample means for the single question on job satisfaction and the ten indices of work experiences, scaled between zero and one. All individual questions and a discussion of these indices are found in Appendix Table 4. Each index is calculated as an average of the scores for the underlying questions, which themselves are drawn from existing scales in the human resources literature, as noted in the methodology section.

Table 4 shows that people with disabilities had a higher score for the turnover intentions index than those without disabilities (0.420 versus 0.323, p<0.01). Examining the specific items in the turnover intentions measure, employees with disabilities have a relatively higher likelihood of planning to look for a job outside of the organization, thinking often of quitting their job, and desiring a new job (Appendix Table 4). These differences, though, were not as large for people who had disclosed their disabilities to their employer.

Table 4. Sample Means for Job Satisfaction and Indices of Work Experiences by Disability Status (Measured as Proportions)

|  | Disability Mean | No Disability Mean | Difference |
|---|---|---|---|
| Somewhat or very satisfied in job | 0.592 | 0.648 | -0.056 |
| Index of agreement on job autonomy | 0.575 | 0.599 | -0.023 |
| Index of turnover intentions | 0.420 | 0.323 | 0.096*** |
| Index of employee organizational commitment | 0.459 | 0.530 | -0.071** |
| Index of employee organizational citizenship behaviors | 0.550 | 0.557 | -0.007 |
| Index of perceived organizational support | 0.309 | 0.403 | -0.094*** |
| Index of employer openness to differences | 0.453 | 0.561 | -0.107*** |
| Index of climate for inclusion | 0.348 | 0.416 | -0.069** |
| Index of treatment of people with disabilities | 0.415 | 0.438 | -0.023 |
| Index of relationship with manager (leader-member exchange) | 0.657 | 0.761 | -0.105*** |
| Index of relationships with coworkers (coworker exchange) | 0.706 | 0.801 | -0.095*** |

Note: Sample size 993. *** statistically significant at 1%, ** at 5%, and * at 10% in 2-tail t tests. Results denote the proportion of respondents who agree with the statements, ranging from 0 to 1. See Appendix Table 4 for sample means using disclosed disability status and for indicators included in the indices.

Closely related, people with disabilities had a lower index of organizational commitment compared to people without disabilities (0.459 versus 0.530, p<0.05). Driving this result was a lower likelihood of people with disabilities to say that they feel a strong sense of belonging at the employer and that they feel like they are a "part of the family." While people with and without disabilities have similar responses to questions about organizational citizenship behaviors, people with disabilities perceive lower levels of support from their organizations (0.309 versus 0.403, p<0.01). Comprising this index of perceived organizational support are perceptions that the employer cares about their well-being and opinions, and that the employer takes pride in their accomplishments at work. These results, however, are muted for people who have disclosed their disabilities to their employer, as shown in Appendix Table 4. In most cases, the disability gap in these job experience indicators is smaller in magnitude and often no longer statistically significant when we focus on people with disclosed disabilities.

Perceptions about the inclusiveness of the workplace mirror this negative relationship between disability status and job satisfaction. Table 4 further shows that people with disabilities are substantially less likely than those without to believe that their employer is open to differences (0.453 versus 0.561, p<0.01). Underlying this index are questions about whether people can reveal their true selves at work, whether employees are valued as people rather than merely for their jobs, and whether the work culture appreciates the differences people bring to the workplace. People with disabilities are also relatively less likely to believe that the employer has an inclusive workplace climate (0.348 versus 0.416, p<0.05), with more skepticism that the employer actively seeks employee input, uses employee insights to redefine work practices, and considers input from people in different roles and functions when problem-solving.

Table 4 also shows that there is no substantial difference between people with and without disabilities in the index of perceptions on how people with disabilities are treated. However, this aggregate index masks some discrepancies between the more detailed questions. People without disabilities tend to have a more favorable view of the culture around disability at their workplace, being more likely to agree that employees treat people with disabilities with respect, and that their manager is responsive to the needs of people with disabilities. As shown in Appendix Table 4, people with disabilities are more likely to state that there is bias against people with disabilities in their workplace, and that employees without disabilities are treated better than employees with disabilities. In many cases the reported differences by disability status are lessened when we consider people with disclosed disabilities. However, there are two instances in which the gap is even larger and highly statistically significant: people with disclosed disabilities are even more likely to state that their workplace has a bias against people with disabilities and that employees without disabilities are treated better than employees with disabilities. In the first case − bias against people with disabilities exists where I work – we see the widest gap, at 0.155 ($p<0.01$), of all the questions in Appendix Table 4.

As shown in Table 4 and Appendix Table 4, people with disabilities are uniformly less likely than people without disabilities to agree to various descriptors of a positive relationship with one's manager, including knowing how satisfied the manager is with one's performance, feeling that the manager is understanding, feeling that the manager recognizes one's potential, feeling that one can count on the manager for support during a tough situation, having an effective working relationship with one's manager, and believing that the manager would use their power and influence to help the employee. These differences by disability status are mirrored in the responses about relationships with one's coworkers. Once we restrict the disability sample to individuals who have disclosed their disabilities to the employer, we see that the experiences of people with disclosed disabilities are closer to those without disabilities in relationships with managers. However, people with disclosed disabilities still have relatively negative views of their relationships with coworkers. This result suggests that the experience of mistreatment/isolation that is more common among people who have disclosed their disabilities is coming more from coworkers than from management. People with disclosed disabilities have particularly low expectations of support and help from coworkers, thus contributing to their workplace alienation and anomie.

The survey results have thus far shown worse perceptions of job experiences among those with disabilities compared to those without, and attenuation of those gaps by disclosure in most indicators, with the notable exceptions of workplace bias, unfavorable treatment, and coworker relations. As shown in Appendix Table 5, which reports regression-adjusted disability gaps, these statistically significant disparities between people with and without disabilities still hold even after controlling for differences in education and other observed characteristics. Consistent with the descriptive analysis, disability status has no statistically significant association with job satisfaction, job autonomy, organizational citizenship behaviors, and perceived treatment of people with disabilities when we control for other demographic characteristics. However, people with a disability have higher turnover intentions (0.095, $p<0.01$), lower organizational commitment (-0.056, $p<0.10$) and perceived organizational support (-0.085, $p<0.05$), and lower scores on their employer's openness to differences (-0.098, $p<0.01$), climate for inclusion (-0.055, $p<0.10$), and relationships with their manager (-0.095, $p<0.01$) and colleagues (-0.067, $p<0.01$).

In the next section, we analyze how an accommodation being granted helps narrow these gaps.

## 6. Accommodations as a moderator in the relationship between disability and job satisfaction

Multivariable regression results for the association between work experiences, disability status, and accommodations are found in Table 5. For job satisfaction and the 10 indicators of workplace experiences, we report results for regressions that include a set of dummy variables for accommodations – modeled as the disposition of accommodation requests – and the interaction of those accommodation variables with disability status. To assess the full difference between employees with and without disabilities in each of the conditions, the regressions with disability interactions do not include a disability main effect, so that the interaction coefficients represent the full disability effect within each condition (e.g., the difference between people with and without disabilities who had an accommodation request fully granted).

Table 5 shows that, not surprisingly, there are adverse effects from having accommodation requests denied or only partially granted on most work experience indicators. The disability interaction coefficients are generally small and insignificant, indicating that accommodation effects are generally similar between people with and without disabilities. Only for turnover intentions do we still see that people with disabilities are more likely to want to leave their jobs compared to people without disabilities (0.096, $p<0.10$), even if they have been fully granted an accommodation request.

Adverse effects of disability on these outcomes are most likely to occur among people with disabilities who have not made an accommodation request: significant adverse effects appear for this group in job satisfaction (-0.303, $p<0.10$), organizational commitment (-0.095, $p<0.10$), perceived organizational support (-0.139, $p<0.05$), employer openness to difference (-0.107, $p<0.10$), climate for inclusion (-0.114, $p<0.05$), manager relations (-0.135, $p<0.01$), and coworker relations (-0.082, $p<0.10$).

Table 5. Regression Results for Association Between Work Experiences, Disability, and Accommodations

| Variable | Job satisfaction | Job autonomy | Turnover intentions | Organizational commitment | Organizational citizenship behaviors | Perceived organizational support |
|---|---|---|---|---|---|---|
| Accom requests (exclude: no request) | | | | | | |
|   Accom request fully granted | -0.041 | -0.067** | 0.026 | -0.035 | 0.005 | 0.005 |
| | (0.115) | (0.032) | (0.036) | (0.037) | (0.034) | (0.036) |
|   Accom request denied | -1.098*** | -0.176*** | 0.346*** | -0.319*** | 0.028 | -0.317*** |
| | (0.135) | (0.038) | (0.042) | (0.043) | (0.040) | (0.042) |
|   Accom request partly granted | -0.745*** | -0.139*** | 0.204*** | -0.259*** | 0.049 | -0.248*** |
| | (0.152) | (0.043) | (0.048) | (0.048) | (0.045) | (0.048) |
| Disability interactions with: | | | | | | |
|   No accom request | -0.303* | -0.037 | 0.060 | -0.095* | -0.001 | -0.139** |
| | (0.178) | (0.050) | (0.056) | (0.057) | (0.053) | (0.056) |
|   Accom request fully granted | -0.182 | 0.074 | 0.096* | 0.007 | 0.062 | -0.038 |
| | (0.171) | (0.048) | (0.054) | (0.054) | (0.051) | (0.054) |
|   Accom request denied | 0.049 | 0.016 | 0.099 | -0.064 | -0.020 | -0.047 |
| | (0.232) | (0.065) | (0.073) | (0.074) | (0.069) | (0.073) |
|   Accom request partly granted | -0.077 | 0.048 | 0.101 | -0.056 | -0.077 | -0.080 |
| | (0.252) | (0.071) | (0.079) | (0.080) | (0.075) | (0.079) |
| Control variables | Yes | Yes | Yes | Yes | Yes | Yes |
| F statistic | 4.88*** | 5.05*** | 7.97*** | 7.72*** | 3.32*** | 8.56*** |
| Multiple R | 0.375 | 0.381 | 0.459 | 0.454 | 0.316 | 0.472 |
| $R^2$ statistic | 0.141 | 0.145 | 0.211 | 0.206 | 0.100 | 0.223 |

Continued on next page.

Table 5 Continued. Regression Results for Association Between Work Experiences, Disability, and Accommodations

| Variable | Employer openness to difference | Climate for inclusion | Treatment of people with disabilities | Manager relations | Coworker relations |
|---|---|---|---|---|---|
| Accom requests (exclude: no request) | | | | | |
|   Accom request fully granted | -0.015 | 0.012 | -0.006 | 0.023 | -0.002 |
| | (0.036) | (0.037) | (0.026) | (0.030) | (0.028) |
|   Accom request denied | -0.321*** | -0.290*** | -0.142*** | -0.247*** | -0.102*** |
| | (0.042) | (0.043) | (0.030) | (0.035) | (0.033) |
|   Accom request partly granted | -0.197*** | -0.206*** | -0.079** | -0.182*** | -0.106*** |
| | (0.048) | (0.049) | (0.035) | (0.040) | (0.038) |
| Disability interactions with: | | | | | |
|   No accom request | -0.107* | -0.114** | -0.038 | -0.135*** | -0.082* |
| | (0.056) | (0.057) | (0.040) | (0.046) | (0.044) |
|   Accom request fully granted | -0.023 | -0.017 | 0.039 | -0.040 | -0.053 |
| | (0.054) | (0.054) | (0.039) | (0.045) | (0.042) |
|   Accom request denied | -0.099 | -0.038 | -0.033 | -0.089 | -0.078 |
| | (0.073) | (0.074) | (0.053) | (0.061) | (0.057) |
|   Accom request partly granted | -0.213*** | -0.041 | -0.050 | -0.129** | -0.030 |
| | (0.079) | (0.080) | (0.057) | (0.066) | (0.062) |
| Control variables | Yes | Yes | Yes | Yes | Yes |
| F statistic | 6.92*** | 5.58*** | 2.97*** | 5.98*** | 2.67*** |
| Multiple R | 0.434 | 0.397 | 0.302 | 0.410 | 0.286 |
| $R^2$ statistic | 0.188 | 0.158 | 0.091 | 0.168 | 0.082 |

Note: Sample size 993. *** statistically significant at 1%, ** at 5%, and * at 10% in 2-tail t tests. All regressions include control variables for age, gender, race/ethnicity, marital status, education, income above $75,000, number of children at home, managerial role, full-time worker, and tenure at the employer.

Among people without disabilities who requested accommodations, having a request denied is associated with reduced job satisfaction (-1.098, $p<0.01$, with an insignificant disability interaction). Relative to people without disabilities who did not make accommodation requests, denial of requests had a significant positive impact on turnover intentions (0.346, $p<0.01$) and an adverse effect on organizational commitment (-0.319, $p<0.01$), perceived organizational support (-0.317, $p<0.01$), employer openness to difference (-0.321, $p<0.01$), climate for inclusion (-0.290, $p<0.01$), treatment of people with disabilities (-0.142, $p<0.01$), manager relations (-0.247, $p<0.01$), and coworker relations (-0.102, $p<0.01$). In each category, the result for being denied an accommodation was statistically significant for people without disabilities compared to those without an accommodation request. The same is true of a request being partly rather than fully granted across categories. Regardless of disability status, having a request fully granted (relative to those with no accommodation request) had no statistically significant effect in any category except job autonomy, where it was negative for those without a disability (-0.067, $p<0.05$).

## 7. Discussion

### 7.1 Interpretation

Overall, we find that the proportion of people who request accommodation is higher among people with a disability compared to people without (H1). Although people with disabilities do not report significantly lower levels of job satisfaction as measured by a single question, those with disabilities do have less favorable work experiences compared to their counterparts without disabilities (H2). Being granted accommodations has a positive moderating effect on the relationship between disability and most measures of work experiences (H3). In this section, we discuss the context and nuances around these findings.

Requests for accommodations do not only come from people with disabilities: we found that over half of people without disabilities had requested accommodations, in addition to almost three-quarters of people with disabilities. A key difference by disability status is that workers with a disability were far more likely to request an accommodation for a health condition, impairment, or disability compared with those without a disability. Even though accommodation requests are quite common, getting those requests fully granted is less common, with about half of *all* requests either denied or only partially granted. Importantly, there is no statistically significant difference between people with and without disabilities in the likelihood of having their requested accommodation granted.

Our results align with an earlier analysis of U.S. microdata across occupations, which found that although a higher proportion of people with disabilities request accommodation compared to people without disabilities, the numerical predominance of people without disabilities in the workforce means that 95% of accommodation requests come from people without disabilities [30]. People with disabilities were significantly more likely to request accommodations than people without disabilities in only four of the 13 industries studied, and 'educational and health services' is one of the four industries. The study also found that people with and without disabilities are equally likely, about 81.5%, to have accommodation requests granted [30].

This seemingly high rate of denying or only partly granting accommodation requests could be explained by Gold et al.'s (2012) analysis of the perspectives of employers, employees, and service providers on the accommodation process [31]. They note that a conflict exists between views of accommodation, particularly between employers' beliefs about the costs of accommodations and employees' opinions that employers focus too much on the cost and legal issues under the ADA. In addition, management practices around accommodations may be unorganized or lack accessibility. In a survey of almost 700 human resource managers, Erickson et al. (2014) found that relatively few employers had a centralized accessibility fund, reviewed the accessibility of their online application system, or evaluated their occupational screenings for bias [32].

Results from our original survey show that people with disabilities have more negative perceptions of their work experiences compared to people without disabilities, including greater turnover intentions, lower sense of organizational support, weaker perceptions of an inclusive workplace, higher perceptions of disability bias, and worse relations with management and coworkers. These relatively worse perceptions can contribute to lower overall job satisfaction for people with disabilities, consistent with earlier studies using national data across occupations [1, 33]. Evidence of lower job satisfaction for people with disabilities can be understood in the context of stigma and identity threat, in which stigma prompts workers with disabilities to adopt self-monitoring behaviors that help maintain their status but may also come with emotional and psychological costs [4].

This pattern was reflected in a set of follow-up interviews we conducted, in which participants without disabilities commented on the "inclusive culture" of the workplace, the willingness of managers to accommodate "reasonable requests," and overall equitable policies and practices. Participants with disabilities, however, were more likely to discuss workplace culture negatively. One interview participant said she felt "pressured and threatened," another said she thought she had to choose between having a needed surgery and keeping her job because she was not allowed to wear a brace while working. Another participant recounted episodes where her co-workers would purposely "test" her to determine if her disability was "real." The same participant said her co-workers would tell new staff, "you can't do this because of her," causing her to feel "singled out" and "uncomfortable."

Some of the biggest and most robust differences between people with and without disabilities in our survey appear in the results for relationships with supervisors and coworkers. A likely explanation is that people are often less attuned to institutional and structural discrimination and much more conscious of the bias they experience in interactions with other people [34]. Hence, the perception of bias in the workplace primarily occurs in the context of relationships with supervisors and coworkers. Of note, many of the disability gaps are muted when we consider people who have disclosed their disability to their employers.

We found that people with disabilities were substantially more likely to request changes to work schedules, moving to another job or location, and requesting a change in communications/information sharing than people without disabilities. Requests for equipment changes were also routine, but we saw no statistically significant differences between people with and without disabilities.

Accommodations at work may be a critical way to moderate some of the negative associations between disability and work experiences. As hypothesized, we found adverse relationships between disability status and many indicators of work experiences, even after including a host of control variables. The adverse relationships between disability and indicators of work experiences (organizational commitment, perceived organizational support, the employer's openness to differences, climate for inclusion, and relationships with the manager and coworkers) are often moderated − albeit not in a statistically significant way − by being fully granted an accommodation. In this context, being granted an accommodation goes beyond ensuring physical accessibility or meeting legal requirements—it involves creating an inclusive workplace culture where employees with disabilities are empowered to engage more fully and succeed [9].

This interpretation of the quantitative results is consistent with responses in the follow-up interviews. Several participants who were fully granted an accommodation spoke highly of the accommodation process. One said, "I had an occupational therapist come out and work with me directly. She made recommendations and everyone was put in place." Another said, "I got a special chair, mouse, and keyboard. I tried them out before choosing—it was a great process." Another said that because of her disability and the distance she had to walk, it was difficult for her to arrive at meetings on time. She said she asked for accommodations and they issued her a formal letter indicating she "would not be penalized for arriving late to meetings."

The quantitative results yield an apparent paradox: the adverse relationship between disability and turnover intentions continues to hold even when an accommodation is fully granted. Several interrelated factors may explain this paradox. First, while accommodations can improve task-related functioning, they do not necessarily address broader issues of workplace climate, such as subtle exclusion, stigma, or lack of psychological safety, which can shape ongoing perceptions of organizational trust and influence future intentions [35]. Second, employees with disabilities may perceive limited opportunities for advancement or a poor fit with organizational culture, which can affect their long-term career decisions regardless of accommodations [36]. Third, external life factors—such as health fluctuations, caregiving responsibilities, or transportation barriers—may disproportionately affect disabled employees' ability to remain in a job, regardless of workplace support [37]. Finally, the drawn-out processes of making a request and providing documentation may help explain why, even when requests were fully granted, employees with disabilities still consider working elsewhere.

Overall, regardless of disability status, people who have a request denied (especially) or less-than-fully granted have statistically significantly more adverse outcomes for all of our job experiences except for one (organizational citizenship behaviors). Denying a request appears to be very bad for the relationship between the employer and the employee, and employers effectively decline to take up the opportunity to improve the experiences of their workers when they deny a request. Having a request denied has similar adverse effects for people with disabilities as for people without disabilities.

The employer in this study has implemented a comprehensive set of formal policies and practices that support disability inclusion for all employees. Among these are a clear non-discrimination policy and an ADA policy outlining procedures for requesting and implementing reasonable accommodations. This ADA policy allows for accommodations not only for individuals with disabilities but also, informally, for employees without a disclosed disability when such accommodations – such as modified schedules, remote work arrangements, or communication supports – promote job performance and work-life balance. In addition, all employees receive annual mandatory education, including training on accessibility, inclusion, and organizational expectations around equity. This inclusive culture, reinforced by HR and leadership, contributes to the high rates of accommodation requests and approvals observed in our data across both groups. As such, the employer's practices may represent a more progressive model of workplace inclusion, which should be considered when assessing the generalizability of findings to other healthcare organizations.

**7.2 Limitations and future directions**

Our study is subject to sampling limitations and potential response bias. In particular, we may underestimate the actual number of individuals with disabilities. The Census questions are designed to capture functional disabilities with more serious limitations. These questions are not intended to replicate the ADA's legal definition of disability, which is broader and includes individuals with a record of, or who are perceived as having, a physical or mental impairment that substantially limits one or more major life activities. Research has shown that these questions may underestimate the population covered by the ADA, particularly those receiving Supplemental Security Income or Social Security Disability Insurance [38]. Moreover, the Census questions may not capture mental health, chronic pain, or people who are neurodivergent and have challenges related to what have historically been called learning disabilities, like ADHD, autism, and a range of other atypicalities. Therefore, our survey includes two additional questions on long-term impairments and challenges in social interactions to capture some of these people. Still, some individuals may hesitate to indicate in a survey that they have difficulty with activities or social interactions due to the persistent stigma surrounding disability.

Further, estimates of accommodation requests for a disability are inherently restricted to those who are aware of or have been diagnosed with a disability, potentially excluding those who remain undiagnosed or unaware. Even individuals with a diagnosed disability who wish to request remote work accommodations may be discouraged by the requirement to provide medical documentation, especially if they are not currently undergoing treatment for their condition.

Another limitation of our study concerns the sampling and recruitment process. Our inability to calculate an accurate response rate introduces uncertainty regarding the representativeness of our sample. Additionally, the relatively high rate of disability disclosure observed in our sample may reflect a self-selection bias, whereby individuals with disabilities or those with a particular interest in disability-related issues were more likely to participate. This potential response bias should be considered when interpreting the findings, as it may limit the generalizability of our results to the broader workforce.

Our survey was limited to the healthcare industry; utilizing a similar instrument to survey a national sample of workers could yield valuable insights on how accommodations can improve workers' experiences on the job. Our results for the healthcare sector indicate that accommodations can be essential for the work experiences of all employees. It would be valuable to explore these topics further across other sectors, with an eye toward expanding opportunities for meaningful employment among people with disabilities and enhancing their workplace experiences once employed.

**8. Summary and practical implications**

This study found a substantial disability gap in various measures of work experience, in that people with disabilities report greater turnover intentions, lower sense of organizational commitment, worse perceptions of organizational support, more negative perceptions about openness and inclusion at work, and worse relations with management and coworkers. Moreover, a surprisingly high proportion of people without disabilities request accommodations (56%, compared to 70% of people with disabilities). Yet, there are no significant differences in the likelihood of having accommodation requests granted. Our regression results show that adverse relationships between disability and perceptions of work experiences are often eliminated when considering the disposition of accommodation requests. People who have been granted accommodations are likelier to report that they are satisfied with their jobs and less likely to report wanting to leave their jobs. Accommodations are also positively associated with employee organizational commitment, perceived organizational support, employer openness to differences, the climate for inclusion, the treatment of people with disabilities, and relationships with their managers.

Our survey of healthcare workers has practical implications for the healthcare sector. As the COVID-19 pandemic has made painfully clear, the economy's health depends on public health and, ultimately, on healthcare workers' health and ability to cope with the demands of their jobs and domestic responsibilities. Public health depends on the healthcare system, which in turn depends on healthcare workers. While healthcare workers may appear to be the primary beneficiaries of accommodations, the benefits are more diffuse [39]. Patients benefit from having healthy physicians and nurses, healthcare workers benefit from their own health, and hospitals benefit from having a healthy workforce. Hence, the healthcare system depends on the health of healthcare workers, which depends on employer policies, including accommodations, to safeguard their health and well-being.

Given the diversity of the sample, our results also have implications for employers across sectors, particularly for the types of accommodations that people request and the positive effects of granting accommodation requests. Accommodations can take many forms, such as shorter work hours, flexible work schedules, lighter workloads, reduced customer or client interaction, postponing more demanding tasks, exchanging duties among colleagues, sharing responsibilities, and allowing additional rest breaks. Modifications to the physical work environment may also be appropriate—for example, using ergonomic equipment, specialized seating or carts, digital organizers, or different lighting options [3]. Employers may also support inclusion by educating managers and supervisors to better understand and respond to disability-related needs, and by fostering a disability-friendly workplace culture.

Prior to the pandemic, employers were generally resistant to granting reasonable accommodations under the ADA [40]. The pandemic may have weakened such resistance by compelling employers to reconsider how best to accomplish job tasks. An important lesson for employers is that accommodations can improve workplace experiences for all employees. Accommodations not only lead to better job satisfaction among accommodated employees, but they can also enhance the engagement of coworkers who generally view the employer more positively when they see a coworker being accommodated [11]. In addition, employer actions to build a more open climate toward people with disabilities can contribute to greater job self-efficacy among employees, allowing employers to fully utilize the work potential of people with disabilities [41]. Granting reasonable accommodations is arguably a vital tool in the toolkit for improving the work potential and job satisfaction of people with disabilities.

Appendix Table 1. Sample Statistics on Disability Status and Disclosure Rates by Disability Type

|  | *Number of respondents with disabilities* | *Percent of respondents with disabilities* |
|---|---|---|
| *Types of Disabilities (Not Mutually Exclusive)* | 228 | 100.0 |
| Deaf/difficulty hearing | 33 | 14.5 |
| Blind/difficulty seeing | 10 | 4.4 |
| Difficulty concentrating/making decisions | 113 | 49.6 |
| Difficulty walking/climbing stairs | 57 | 25.0 |
| Difficulty dressing/bathing | 7 | 3.1 |
| Difficulty doing errands alone | 39 | 17.2 |
| Difficulty interacting with others | 47 | 20.6 |
| Long-term health impairment | 106 | 46.5 |
|  |  |  |
| *Disclosure Among People with Disabilities* | 228 | 100.0 |
| Have you disclosed your health condition, impairment, or disability to your employer? |  |  |
|   Yes | 117 | 51.8 |
|   No | 57 | 25.2 |
|   It's complicated | 52 | 23.0 |
|   Did not respond | 2 | 0.9 |
|  |  |  |
| *Disclosure Rates by Disability Type* |  |  |
|   Yes | 117 | 51.8 |
|     Deaf/difficulty hearing | 19 | 57.6 |
|     Blind/difficulty seeing | 4 | 40.0 |
|     Difficulty concentrating/making decisions | 46 | 40.7 |
|     Difficulty walking/climbing stairs | 39 | 68.4 |
|     Difficulty dressing/bathing | 3 | 42.9 |
|     Difficulty doing errands alone | 19 | 48.7 |
|     Difficulty interacting with others | 19 | 40.4 |
|     Long-term health impairment | 65 | 61.3 |
|   No | 57 | 25.2 |
|     Deaf/difficulty hearing | 10 | 30.3 |
|     Blind/difficulty seeing | 4 | 40.0 |
|     Difficulty concentrating/making decisions | 32 | 28.3 |
|     Difficulty walking/climbing stairs | 8 | 14.0 |
|     Difficulty dressing/bathing | 4 | 57.1 |
|     Difficulty doing errands alone | 11 | 28.2 |
|     Difficulty interacting with others | 15 | 31.9 |
|     Long-term health impairment | 17 | 16.0 |
|   It's complicated | 52 | 23.0 |
|     Deaf/difficulty hearing | 4 | 12.1 |
|     Blind/difficulty seeing | 2 | 20.0 |
|     Difficulty concentrating/making decisions | 34 | 30.1 |
| Difficulty walking/climbing stairs | 9 | 15.8 |
| Difficulty dressing/bathing | 0 | 0.0 |
| Difficulty doing errands alone | 8 | 20.5 |
| Difficulty interacting with others | 13 | 27.7 |
| Long-term health impairment | 23 | 21.7 |
| Did not respond | 2 | 0.9 |

Note: Sample size: 228 persons with a disability. Types of disability are not mutually exclusive.

Appendix Table 2. Sample Means for Demographic Characteristics by Disclosed Disability Status (Measured as Proportions Unless Indicated Otherwise)

| Demographic Characteristic | Disclosed Disability Mean | Undisclosed/ No Disability Mean | Difference |
|---|---|---|---|
| Age | 45.468 | 44.530 | 0.938 |
| # Children at home | 1.809 | 2.010 | -0.202* |
| Gender | | | |
|   Man | 0.079 | 0.106 | -0.027 |
|   Woman | 0.877 | 0.861 | 0.016 |
|   Nonbinary | 0.044 | 0.032 | 0.011 |
| Race/Ethnicity | | | |
|   Black | 0.112 | 0.120 | -0.008 |
|   White | 0.871 | 0.832 | 0.038 |
|   Hispanic | 0.052 | 0.027 | 0.025 |
|   American Indian/Alaska Native | 0.026 | 0.042 | -0.016 |
|   Asian/Pacific Islander | 0.009 | 0.017 | -0.009 |
|   Multiracial/other | 0.060 | 0.060 | 0.000 |
| Married | 0.482 | 0.643 | -0.161*** |
| Education | | | |
|   <9th grade | 0.000 | 0.001 | -0.001 |
|   High school graduate | 0.121 | 0.069 | 0.051** |
|   Some college, no degree | 0.207 | 0.155 | 0.052 |
|   Associate degree | 0.155 | 0.178 | -0.023 |
|   Bachelor's degree | 0.336 | 0.360 | -0.024 |
|   Master's degree | 0.147 | 0.178 | -0.032 |
|   Professional degree/PhD | 0.034 | 0.058 | -0.023 |
| Income >=$75,000 | 0.216 | 0.348 | -0.132*** |
| Works full-time | 0.872 | 0.885 | -0.013 |
| Worked at employer before pandemic | 0.726 | 0.794 | -0.068* |
| Worked at employer <= 5 years | 0.547 | 0.410 | 0.137*** |
| Occupation | | | |
|   Administrative support staff | 0.216 | 0.176 | 0.040 |
|   Professional/technical staff | 0.138 | 0.131 | 0.007 |
|   Nurse | 0.198 | 0.255 | -0.056 |
|   Physician | 0.000 | 0.007 | -0.007 |
|   Healthcare aide | 0.078 | 0.046 | 0.032 |
|   Other healthcare provider | 0.233 | 0.210 | 0.023 |
|   Manager | 0.086 | 0.121 | -0.035 |
|   Other service provider | 0.052 | 0.055 | -0.003 |

Note: This Appendix Table reports sample means for the disability gap when the disability sample is narrowed to those who disclosed their disability to their employer. Sample size 993. *** statistically significant at 1%, ** at 5%, and * at 10% in 2-tail t tests. Results denote the proportion of respondents who gave the indicated response in the survey, ranging from 0 to 1.

Appendix Table 3. Sample Means for Accommodation Requests by Disclosed Disability Status (Measured as Proportions)

| Accommodations | Disclosed Disability Mean | Undisclosed/ No Disability Mean | Difference |
|---|---|---|---|
| Have you ever requested accommodations? | 0.730 | 0.574 | 0.156*** |
|   Type requested: equipment | 0.452 | 0.476 | -0.024 |
|   Type requested: physical change to workplace | 0.333 | 0.256 | 0.077 |
|   Type requested: work from home | 0.298 | 0.278 | 0.020 |
|   Type requested: change to work schedule | 0.679 | 0.566 | 0.113* |
|   Type requested: restructure job | 0.262 | 0.198 | 0.064 |
|   Type requested: move to another job or location | 0.262 | 0.142 | 0.120*** |
|   Type requested: change communications/info sharing | 0.560 | 0.454 | 0.106* |
|   Type requested: other | 0.274 | 0.184 | 0.090* |
| Most recent request: equipment | 0.151 | 0.211 | -0.060 |
| Most recent request: physical change to workplace | 0.058 | 0.058 | 0.000 |
| Most recent request: work from home | 0.070 | 0.066 | 0.004 |
| Most recent request: change to work schedule | 0.302 | 0.281 | 0.021 |
| Most recent request: restructure job | 0.116 | 0.076 | 0.041 |
| Most recent request: move to another job or location | 0.081 | 0.058 | 0.024 |
| Most recent request: change communications/info sharing | 0.140 | 0.179 | -0.040 |
| Most recent request: other | 0.081 | 0.072 | 0.010 |
| Most recent accommodation was requested within the past 12 months | 0.786 | 0.826 | -0.040 |
| Did you request this change in order to accommodate any health condition, impairment, or disability that you may have? | 0.393 | 0.104 | 0.289*** |
| Was the requested change or accommodation made? | | | |
|   Yes | 0.512 | 0.477 | 0.035 |
|   No | 0.262 | 0.297 | -0.035 |
|   Partially | 0.226 | 0.226 | 0.000 |

Note: This Appendix Table reports sample means for the disability gap when the disability sample is narrowed to those who disclosed their disability to their employer. Results denote the proportion of respondents who gave the indicated response in the survey, ranging from 0 to 1. Sample size 993 for first question. Responses for remaining questions are conditional on having ever requested accommodations. *** statistically significant at 1%, ** at 5%, and * at 10% in 2-tail t tests.

Appendix Table 4. Detailed Indicators and Sample Means for Job Experiences by Disability Status (Measured as Proportions)

| % of People Who Agree with These Statements: | Self-Reported Disb. Mean | Self-Reported No Disb. Mean | Self-Reported Disb. Diff. | Disclosed Disb. Mean | Disclosed NoDisb/Undisc Mean | Disclosed Disb. Diff. |
|---|---|---|---|---|---|---|
| **Somewhat or very satisfied in job** | **0.592** | **0.648** | **-0.056** | **0.590** | **0.641** | **-0.051** |
| **Index of agreement on job autonomy** | **0.575** | **0.599** | **-0.023** | **0.593** | **0.593** | **0.001** |
| Currently pandemic job allows me to decide when to begin and end work each day | 0.356 | 0.391 | -0.035 | 0.365 | 0.385 | -0.020 |
| Currently I have some control over the sequencing of my work activities | 0.704 | 0.744 | -0.040 | 0.707 | 0.738 | -0.031 |
| Currently I can decide when to do particular work activities | 0.665 | 0.665 | 0.000 | 0.707 | 0.659 | 0.048 |
| **Index of turnover intentions** | **0.420** | **0.323** | **0.096***** | **0.407** | **0.337** | **0.070*** |
| Currently I plan to look for job outside this company during the next year | 0.363 | 0.283 | 0.080** | 0.359 | 0.294 | 0.065 |
| Currently I often think about quitting my job at this company | 0.439 | 0.334 | 0.104*** | 0.419 | 0.350 | 0.069 |
| Currently I want to get a new job | 0.461 | 0.353 | 0.108*** | 0.444 | 0.368 | 0.076 |
| **Index of organizational commitment** | **0.459** | **0.530** | **-0.071**** | **0.481** | **0.518** | **-0.037** |
| I feel a strong sense of 'belonging' to the employer | 0.469 | 0.577 | -0.108*** | 0.470 | 0.563 | -0.093* |
| I feel like 'part of the family' at the employer | 0.427 | 0.526 | -0.099*** | 0.436 | 0.513 | -0.077 |
| The employer has a great deal of personal meaning for me | 0.476 | 0.487 | -0.011 | 0.538 | 0.477 | 0.061 |
| **Index of organizational citizenship behaviors** | **0.550** | **0.557** | **-0.007** | **0.584** | **0.551** | **0.033** |
| I keep up with developments at the employer | 0.610 | 0.634 | -0.024 | 0.615 | 0.630 | -0.014 |
| I offer ideas to improve the functioning of the employer | 0.430 | 0.450 | -0.020 | 0.453 | 0.444 | 0.009 |
| I take action to protect the employer from potential problems | 0.610 | 0.588 | 0.022 | 0.684 | 0.581 | 0.103** |
| **Index of perceived organizational support** | **0.309** | **0.403** | **-0.094**** | **0.376** | **0.383** | **-0.006** |
| The employer really cares about my well-being | 0.339 | 0.433 | -0.093** | 0.371 | 0.417 | -0.046 |
| The employer takes pride in my accomplishments at work | 0.313 | 0.412 | -0.100*** | 0.405 | 0.387 | 0.018 |
| The employer cares about my opinions | 0.276 | 0.366 | -0.090** | 0.353 | 0.344 | 0.009 |

Appendix Table 4 Continued. Detailed Indicators and Sample Means for Job Experiences, by Disability Status

|  | Self-Reported | | | Disclosed | | |
|---|---|---|---|---|---|---|
| *% of People Who Agree with These Statements:* | Disb. Mean | No Disb. Mean | Disb. Diff. | Disb. Mean | NoDisb/Undisc Mean | Disb. Diff. |
| **Index of employer openness to differences** | **0.453** | **0.561** | **-0.107***** | **0.490** | **0.542** | **-0.052** |
| The employer has a non-threatening environment in which people can reveal their 'true' selves | 0.482 | 0.616 | -0.133*** | 0.521 | 0.594 | -0.072 |
| Employees are valued for who they are as people, not just for the jobs that they fill | 0.355 | 0.450 | -0.094** | 0.402 | 0.432 | -0.030 |
| We have a culture in which employees appreciate the differences that people bring to the workplace | 0.522 | 0.616 | -0.095** | 0.547 | 0.601 | -0.054 |
| **Index of climate for inclusion** | **0.348** | **0.416** | **-0.069**** | **0.385** | **0.403** | **-0.018** |
| Employee input is actively sought | 0.390 | 0.463 | -0.072* | 0.402 | 0.452 | -0.050 |
| Everyone's opinions for how to do things better are given serious consideration | 0.298 | 0.343 | -0.045 | 0.333 | 0.333 | 0.001 |
| Employees' insights are used to rethink or redefine work practices | 0.325 | 0.399 | -0.075** | 0.359 | 0.385 | -0.026 |
| Management exercises the belief that problem-solving is improved when input from different roles, ranks, and functions is considered | 0.377 | 0.463 | -0.086** | 0.444 | 0.443 | 0.001 |
| **Index of treatment of people with disabilities** | **0.415** | **0.438** | **-0.023** | **0.459** | **0.429** | **0.030** |
| Employees with disabilities have the same opportunities as people without disabilities | 0.469 | 0.496 | -0.027 | 0.521 | 0.486 | 0.036 |
| The employer is making strong efforts to improve conditions and opportunities for people with disabilities | 0.382 | 0.439 | -0.058 | 0.419 | 0.427 | -0.008 |
| The environment is such that, when accommodations are made, people with disabilities can be just as productive as people without disabilities | 0.447 | 0.489 | -0.041 | 0.513 | 0.475 | 0.038 |
| Bias against people with disabilities exists where I work | 0.259 | 0.122 | 0.136*** | 0.291 | 0.135 | 0.155*** |
| Where I work, employees without disabilities are treated better than employees with disabilities | 0.137 | 0.067 | 0.070*** | 0.162 | 0.072 | 0.090*** |
| Top management commits to hire people with disabilities | 0.138 | 0.174 | -0.036 | 0.148 | 0.168 | -0.021 |
| Employees treat people with disabilities with respect | 0.586 | 0.679 | -0.093*** | 0.624 | 0.662 | -0.039 |
| My manager treats people with disabilities with respect | 0.659 | 0.702 | -0.042 | 0.707 | 0.690 | 0.017 |
| The employer is responsive to the needs of people with disabilities | 0.471 | 0.532 | -0.061 | 0.513 | 0.519 | -0.006 |
| My manager is responsive to the needs of people with disabilities | 0.608 | 0.672 | -0.064* | 0.692 | 0.652 | 0.040 |

Appendix Table 4 Continued. Detailed Indicators and Sample Means for Job Experiences, by Disability Status

| % of People Who Agree with These Statements: | Self-Reported | | | Disclosed | | |
|---|---|---|---|---|---|---|
| | Disb. Mean | No Disb. Mean | Disb. Diff. | Disb. Mean | NoDisb/Undisc Mean | Disb. Diff. |
| **Index of manager relations (leader-member exchange)** | **0.657** | **0.761** | **-0.105***** | **0.706** | **0.742** | **-0.036** |
| I usually know how satisfied my manager is with what I do | 0.699 | 0.801 | -0.101*** | 0.726 | 0.784 | -0.058 |
| I feel that my manager understands my problems and needs | 0.562 | 0.712 | -0.150*** | 0.615 | 0.686 | -0.070 |
| I feel that my manager recognizes my potential | 0.611 | 0.750 | -0.140*** | 0.650 | 0.728 | -0.078* |
| I can count on my manager to support me even when I'm in a tough situation at work | 0.668 | 0.755 | -0.086*** | 0.718 | 0.737 | -0.019 |
| I would support my manager's decision even if he or she was not present | 0.752 | 0.801 | -0.049 | 0.821 | 0.786 | 0.035 |
| I have an effective working relationship with my manager | 0.730 | 0.821 | -0.091*** | 0.769 | 0.804 | -0.035 |
| If necessary, my manager would use his or her power and influence to help me | 0.575 | 0.692 | -0.117*** | 0.641 | 0.669 | -0.028 |
| **Index of coworker relations (coworker exchange)** | **0.706** | **0.801** | **-0.095***** | **0.698** | **0.790** | **-0.092***** |
| I usually know how satisfied my coworkers are with what I do | 0.681 | 0.804 | -0.123*** | 0.672 | 0.790 | -0.117*** |
| I feel that my coworkers understand my problems and needs | 0.590 | 0.711 | -0.121*** | 0.595 | 0.695 | -0.100** |
| I can count on my coworkers to support me even when I'm in a tough situation at work | 0.740 | 0.818 | -0.078*** | 0.724 | 0.811 | -0.086** |
| I would support my coworkers' decisions even if they were not present | 0.811 | 0.837 | -0.027 | 0.810 | 0.834 | -0.024 |
| I have an effective working relationship with my coworkers | 0.828 | 0.893 | -0.064*** | 0.802 | 0.888 | -0.086*** |
| If necessary, my coworkers would use their power and influence to help me | 0.586 | 0.744 | -0.158*** | 0.586 | 0.724 | -0.138*** |

Note: This Appendix Table reports sample means for the disability gap when using the sample of people who reported a disability in the survey (columns 1-3), and the smaller sample who disclosed their disability to their employer (columns 4-6). Sample size 993. *** statistically significant at 1%, ** at 5%, and * at 10% in 2-tail t tests. Results denote the proportion of respondents who agree with the statements, ranging from 0 to 1. Indicators in each index are weighted equally.

Appendix Table 5. Regression-Adjusted Disability Gaps in Perceptions of Work Experiences

|  | Regression-Adjusted Disability Gap |
|---|---|
| Somewhat or very satisfied in job | -0.047 |
|  | (0.038) |
| Index of agreement on job autonomy | 0.015 |
|  | (0.028) |
| Index of turnover intentions | 0.095*** |
|  | (0.033) |
| Index of employee organizational commitment | -0.056* |
|  | (0.033) |
| Index of employee organizational citizenship behaviors | 0.015 |
|  | (0.030) |
| Index of perceived organizational support | -0.085** |
|  | (0.033) |
| Index of employer openness to differences | -0.098*** |
|  | (0.033) |
| Index of climate for inclusion | -0.055* |
|  | (0.033) |
| Index of treatment of people with disabilities | -0.013 |
|  | (0.023) |
| Index of relationship with manager (leader-member exchange) | -0.095*** |
|  | (0.027) |
| Index of relationships with coworkers (coworker exchange) | -0.067*** |
|  | (0.025) |

Note: Sample size 993. *** statistically significant at 1%, ** at 5%, and * at 10% in 2-tail t tests. Coefficients above are from separate estimations of each work experience measure regressed on a dummy variable for disability plus control variables for age, gender, race/ethnicity, marital status, education, income above $75,000, number of children at home, managerial role, full-time worker, and tenure at the employer.

# Appendix Notes:

# Sample and Perceptions of Work Experience Scales

**Sample:**

Our survey instrument included questions on employees' awareness and perceptions of employer policies that address the physical and mental health needs of workers. It focused on employer practices around work from home, and it also asked about work experiences before the pandemic. Survey questions used the wording "Before March 2020" to denote the period before the pandemic started, and phrases such as "currently" or "today" to denote the current period at the time of the survey. We used the data to calculate simple summary statistics on the prevalence of work from home, disability disclosure, perceptions of workplace inclusiveness, treatment of people with disabilities, and various measures of job satisfaction.

In collaboration with partners from the healthcare system, our research team distributed the survey link via Qualtrics to employees. The distribution included a cover letter providing details about the study and outlining the informed consent process.

Qualtrics reported 1,405 respondents. We dropped 135 of those respondents because they clicked on the survey link but did not answer any questions. An additional 277 respondents did not respond to questions about their disability status, so we also dropped these individuals, leaving a sample of 993. Robustness checks in which these 277 respondents were kept in the sample and assumed to have no disability yielded substantively similar results to those reported in the paper.

**Perceptions of Work Experience Scales:**

Perceived organizational support:

    a. The organization really cares about my well-being.
    b. The organization takes pride in my accomplishments at work.
    c. The organization cares about my opinions.

Cites:

3 items from Wayne et al. based on longer scale from Eisenberger et al.

Wayne, S., Shore, L., & Liden, R. (1997). Perceived organizational support and leader-member exchange: A social exchange perspective. *Academy of management Journal, 40,* 82-111.

Eisenberger, R., Huntington, R., Hutchison, S., & Sowa, D. (1986). Perceived organizational support. *Journal of Applied Psychology, 71,* 500-507.

Organizational commitment:

This is known more specifically as "affective organizational commitment"

a. I feel a strong sense of "belonging" to my organization.
b. I feel like "part of the family" at my organization.
c. My organization has a great deal of personal meaning for me.

Cite:

Drawn from Meyer, J.P, Allen, M.J., & Smith, C.A. (1993). Commitment to organizations and occupations: Extension and test of a three-component conceptualization. *Journal of Applied Psychology, 78,* 538-551.

Organizational citizenship behaviors:

How often do you engage in these behaviors

a. Keep up with developments in the organization.
b. Offer ideas to improve the functioning of the organization.
c. Take action to protect the organization from potential problems.

Cite:

Lee, K. and Allen, N.J. (2002), "Organizational citizenship behavior and workplace deviance: the role of affect and cognitions", Journal of Applied Psychology, Vol. 87 No. 1, pp. 131-142, doi: 10.1037// 0021-9010.87.1.131.

Leader-member exchange:

Please indicate the extent to which you agree with these statements about the relationship between you and your supervisor/manager? *(Please circle ONE answer for each item)*

I usually know how satisfied my manager is with what I do.

I feel that my manager understands my problems and needs

I feel that my manager recognizes my potential.

I can count on my manager to support me even when I'm in a tough situation at work.

I would support my manager's decisions even if he or she was not present.

I have an effective working relationship with my manager.

If necessary, my manager would use his or her power and influence to help me.

Adapted from:

Graen, G. B., & Uhl-Bien, M. (1995). Relationship-based approach to leadership: Development of leader-member exchange (LMX) theory of leadership over 25 years: Applying a multi-level multi-domain perspective. *The leadership quarterly*, *6*(2), 219-247.

Coworker exchange:

Please indicate the extent to which you agree with these statements about the relationship between you and your coworkers (Please circle ONE answer for each item) ※

E34. I usually know how satisfied my coworkers are with what I do.

E35. I feel that my coworkers understand my problems and needs.

E36. I can count on my coworkers to support me even when I'm in a tough situation at work.

E37. I would support my coworkers' decisions even if they were not present.

E38. I have an effective working relationship with my coworkers.

E39. If necessary, my coworkers would use their power and influence to help me.

Cite:

Sherony, K. M., & Green, S. G. (2002). Coworker exchange: relationships between coworkers, leader-member exchange, and work attitudes. *Journal of applied psychology*, *87*(3), 542.

Climate for inclusion:

This is a subscale representing the dimension of climate for inclusion in decision-making

How would you rate the inclusiveness of your organization with regard to employees' ideas and experiences in general, particularly on each of the following? *Please CHOOSE ONE answer for each item.*

- E13. In my organization, employee input is actively sought.
- E14. In my organization, everyone's ideas for how to do things better are given serious consideration.
- E15. In my organization, employees' insights are used to rethink or redefine work practices.
- E16. Management exercises the belief that problem-solving is improved when input from different roles, ranks, and functions is considered.

Cite:

Nishii, L. H. (2013). The benefits of climate for inclusion for gender-diverse groups. *Academy of Management journal*, *56*(6), 1754-1774.

Turnover intentions:

C9. I will look for a job outside this company during the next year.

C10. I often think about quitting my job at this company.

C11. I would like to get a new job.

Cite:

Konovsky, M. A., & Cropanzano, R. (1991). Perceived fairness of employee drug testing as a predictor of employee attitudes and job performance. *Journal of applied psychology*, *76*(5), 698.

Autonomy:

C5. The job denies me any chance to use my personal initiative or judgment in carrying out the work.

C6. The job gives me considerable opportunity for independence and freedom in how I do the work.

C7. The job gives me considerable flexibility to work at my personal "peak" times (i.e., the times of day I feel most productive)

C8. I have complete freedom to schedule my own work hours.

These four are all from Desroisers (2001), drawing from Hackman and Oldham (1978) and Breaugh (1985).

Cites:

Breaugh, J. A. (1985). The measurement of work autonomy. *Human relations*, *38*(6), 551-570.

Desrosiers, E. I. (2001). *Telework and work attitudes: The relationship between telecommuting and employee job satisfaction, organizational commitment, perceived organizational support, and perceived co-worker support*. Purdue University.

Hackman, J. R., & Oldham, G. R. (1975). Development of the job diagnostic survey. *Journal of Applied psychology*, *60*(2), 159.

The accommodations questions are all from:

Schur, L., Nishii, L., Adya, M., Kruse, D., Bruyère, S. M., & Blanck, P. (2014). Accommodating employees with and without disabilities. *Human Resource Management*, *53*(4), 593-621.